\documentclass[aps,pre,10pt,twocolumn,superscriptaddress,showpacs]{revtex4-1}
\usepackage{graphicx}
\usepackage{color}

\begin{document}

\title{Efficient determination of alloy ground-state structures}
\author{Atsuto \surname{Seko}}
\email{seko@cms.mtl.kyoto-u.ac.jp}
%\homepage{http://cms.mtl.kyoto-u.ac.jp/seko-e.html}
\affiliation{Department of Materials Science and Engineering, Kyoto University, Kyoto 606-8501, Japan}
\affiliation{Center for Elements Strategy Initiative for Structure Materials (ESISM), Kyoto University, Kyoto 606-8501, Japan}
\author{Kazuki \surname{Shitara}}
\affiliation{Department of Materials Science and Engineering, Kyoto University, Kyoto 606-8501, Japan}
\author{Isao \surname{Tanaka}}
\affiliation{Department of Materials Science and Engineering, Kyoto University, Kyoto 606-8501, Japan}
\affiliation{Center for Elements Strategy Initiative for Structure Materials (ESISM), Kyoto University, Kyoto 606-8501, Japan}
\affiliation{Nanostructures Research Laboratory, Japan Fine Ceramics Center, Nagoya 456-8587, Japan}

\date{\today}


\begin{abstract}
We propose an efficient approach to accurately finding the ground-state structures in alloys based on the cluster expansion method.
In this approach, a small number of candidate ground-state structures are obtained without any information of the energy.
To generate the candidates, we employ the convex hull constructed from the correlation functions of all possible structures by using an efficient algorithm.
This approach is applicable to not only simple lattices but also complex lattices.
Firstly, we evaluate the convex hulls for binary alloys with four types of simple lattice. 
Then we discuss the structures on the vertices.
To examine the accuracy of this approach, we perform a set of density functional theory calculations and the cluster expansion for Ag-Au alloy and compare the formation energies of the vertex structures with those of all possible structures.
As applications, the ground-state structures of the intermetallic compounds CuAu, CuAg, CuPd, AuAg, AuPd, AgPd, MoTa, MoW and TaW are similarly evaluated.
Finally, the energy distribution is obtained for different cation arrangements in MgAl$_2$O$_4$ spinel, for which long-range interactions are essential for the accurate description of its energetics.
\end{abstract}

\maketitle

\section{Introduction}
Knowledge of the crystal structure is essential for understanding the physical properties of solids. 
The determination of yet-unknown structures has therefore been an important issue in physics and materials science.
Many techniques have been developed recently to explore the most stable crystal structure only from information of the constituent elements using density functional theory (DFT) calculations\cite{PhysRev.136.B864,PhysRev.140.A1133}.
They can be roughly categorized into two groups.
One is to use a global optimization algorithm starting from an initial configuration\cite{oganov2011modern,trimarchi2008finding,oganov2006crystal,woodley2008crystal,pickard2011ab,PhysRevB.82.094116,PhysRevB.87.184104,PSSB:PSSB200945246,Laio01102002}.
The ground-state structure with a fixed composition is expected to be found without any prior knowledge of the crystal structure.
The other method uses a crystal structure database. 
The ground-state structures are searched for among structures included in the database with the aid of a machine learning technique, which enables the efficient search for ground-state structures\cite{curtarolo2003predicting,fischer2006predicting}.
The phase stability among a large number of structure types can be comparatively discussed. 
%although it can be evaluated only from within observed structures.
%Recently, several databases are available, which contain formation energies of a large number of alloys and compounds and phase diagrams since it has been possible to perform an astronomical number of DFT calculations[cite,MatGenome,cartaloro,OQMD(wolverton)].

These techniques have also been applied to determine the ground-state structures in alloys.
A combination of DFT calculation and the cluster expansion (CE) method\cite{CE1,CE2,CE3} is also useful for alloys.
Recent progress in combining the CE method with DFT calculations\cite{CV2,geneAlgo1,Jansen_2008_Bayesian,Mueller_Ceder_2009_Bayesian,Mueller_2010_structure_selection,DiazOrtiz_2010,sampling:seko,casp:seko,nelson2013compressive,nelson2013cluster,kristensen2013relative,sanchez2010cluster} has enabled us to evaluate the ground-state structures and phase stability accurately.
Although it is impossible to find structures beyond a given crystal lattice using the ordinary CE method, many yet-unobserved structures have been discovered within alloy configurations on the crystal lattice.
%The combination of the database approach and the CE method has also been adopted to overcome the weaknesses of both the database approach and CE method\cite{PhysRevB.81.024112,}.

An alternative approach to finding alloy ground-state structures is to analyze the correlation functions defined in the CE method\cite{Hart_Nature2,Fujimura_2013_AENM:AENM201300060,Shitara-Bi2O3}.
This approach is based on the hypothesis that the energy shows an extremum (maximum or minimum) in the ``least random'' structure that has a high relative likelihood index on the basis of the inspection of existing structures of binary compounds.
The likelihood index is given as the sum of the squares of correlation functions. 
According to the likelihood index, a small number of structures can be chosen for the DFT calculation from a large number of candidate structures without any information of the energy. 
By computing the energies of only the chosen structures, the ground-state structures can be efficiently estimated.

In this study, a more elegant approach to efficiently determining the ground-state structures is demonstrated.
To obtain a small number of candidate ground-state structures without any information of the energy, we use the convex hull defined in the space of the correlation functions containing all possible structures. 
This is based on the widely accepted knowledge that the alloy ground-state structures correspond to part of the structures on the vertices of the configurational polyhedron\cite{CE3,kudo1976method}. 
As an approximation, here we assume that the configurational polyhedron is the same as the convex hull estimated from the correlation functions of all possible structures.
The convex hull is obtained using an efficient numerical algorithm.
The use of the numerical algorithm enables us to obtain the convex hull easily for not only simple lattices but also complex lattices.

This study is organized as follows.
Firstly, we evaluate the convex hulls for four types of simple lattice, i.e., face-centered cubic (fcc), body-centered cubic (bcc), hexagonal close-packed (hcp) and simple cubic (sc) lattices. 
In addition, the structures on the vertices of the convex hulls are carefully examined. 
We then carry out the CE method for fcc Ag-Au alloy to examine the accuracy of our approach. 
Since the energies for a large number of structures can be quickly computed by the CE method, the computed formation energies of all structures are compared with those of the structures on vertices.
As applications, we evaluate the ground-state structures of nine intermetallic compounds, i.e., CuAu, CuAg, CuPd, AuAg, AuPd, AgPd, MoTa, MoW and TaW.
Finally, we extend this approach to estimate the energy distribution of MgAl$_2$O$_4$ spinel, where the long-range electrostatic interactions are essential in describing its energetics accurately.

\section{Methodology}
\begin{figure}[tbp]
\begin{center}
\includegraphics[width=0.8\linewidth]{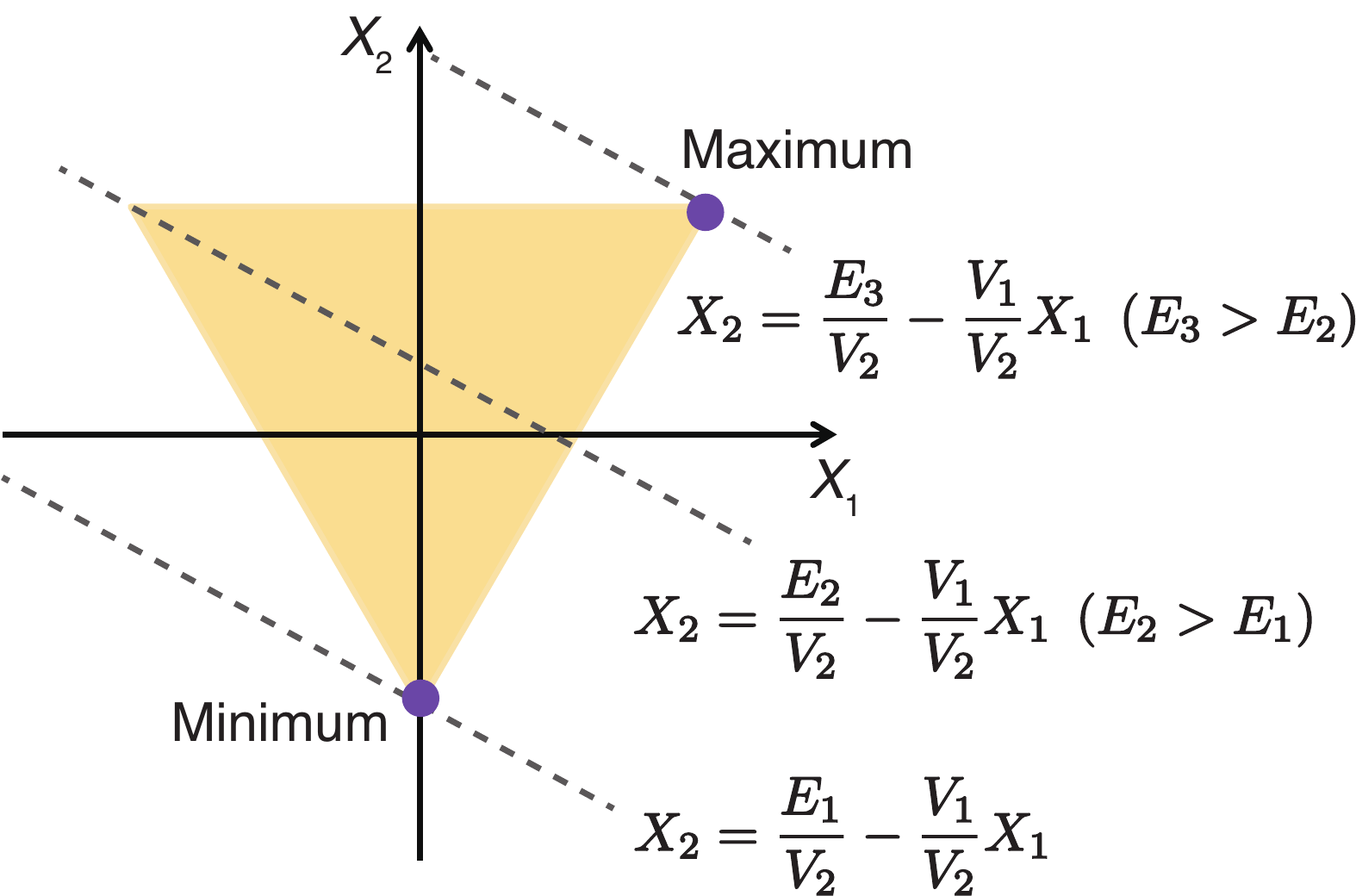} 
\caption{
Schematic illustration of the configurational polyhedron (triangle) in a two-dimensional correlation function space.
The correlation functions of two clusters ($X_1$ and $X_2$) are bounded by the triangle.
The energy and correlation functions ($X_1$ and $X_2$) have a linear relationship described as $ E = V_1 X_1 + V_2 X_2$. 
Constant-energy surfaces are expressed by dotted lines.
Only the ratio $V_1/V_2$ determines the ground-state structure.
}
\label{special:schematic_polyhedron}
\end{center}
\end{figure}

The efficient approach to estimating the ground-state structures is based on the CE method.
Within the formalism of the CE method for binary alloys, an alloy configuration is expressed only by a set of correlation functions.
The correlation function of cluster $\alpha$, $ \varphi_\alpha $, is described using the pseudospin configurational variable $\sigma_i$ for the respective lattice site $i$ as
\begin{equation}
\varphi_\alpha = \frac{1}{N_\alpha} \sum\limits_{i,j,\cdots} \sigma_i \sigma_j \cdots,
\label{special:correlation_function}
\end{equation}
where $N_\alpha$ denotes the number of cluster $\alpha$ included in a structure.
The summation is taken over all clusters included in the structure.
Since the pseudospin variables are commonly set to $+1$ and $-1$ for binary alloys, the correlation functions range from $-1$ to $+1$ when there is no constraint on the correlation functions.
However, the ranges are actually restricted to with a polyhedron determined only by a given crystal lattice.
This is called the configurational polyhedron.
Figure \ref{special:schematic_polyhedron} schematically illustrates a two-dimensional configurational polyhedron (triangle) in a correlation function space.
As illustrated in Fig. \ref{special:schematic_polyhedron}, the range of the correlation functions is located inside the configurational polyhedron.
A detailed analysis of the configurational polyhedra can be found in the book of Ducastelle\cite{CE3}.

Our purpose is to find the ground-state structures among the structures inside the configurational polyhedron.
In the CE method for binary alloys, the energy $E$ for an alloy configuration has a linear relationship with the correlation functions, expressed as
\begin{eqnarray}
E = \sum\limits_{\alpha} V_\alpha \cdot \varphi_\alpha, 
\label{hamiltonian}
\end{eqnarray}
where $V_\alpha$ is called the effective cluster interaction (ECI) of cluster $\alpha$.
Once the ECIs are given, a constant-energy surface is expressed by a straight line in a two-dimensional configurational space as shown in Fig. \ref{special:schematic_polyhedron}. 
Therefore, the structures with the minimum and maximum energies correspond to those on vertices.
We call these structures ``vertex structures''.
To find the ground-state structures, there is no need to consider any structures other than the vertex structures.

In the literature, analytical configurational polyhedra were derived for simple lattices with a small number of interactions\cite{CE3,kanamori1966,kaburagi1975method,kanamori1977conditions,lipkin1988ground,PhysRevB.27.3018,richards1971pairwise,allen1972ground,allen1973ground,cahn1979ground,sanchez1981structure,kudo1976method}.
However, it is generally difficult to obtain an $n$-dimensional polyhedron analytically.
Therefore, we here approximate an $n$-dimensional configurational polyhedron as the convex hull of the correlation functions, numerically estimated from a large number of possible alloy configurations.
The convex hull is searched for using the quickhull algorithm\cite{barber1996quickhull}, which is a method of computing the convex hull of a finite set of points in a given space.
The set of possible alloy configurations is prepared by the derivative structure search\cite{Hart_derivativestructure,Hart_derivativestructure2}.
Consequently, the accuracy of this approach to finding the ground-state structures depends on the number of clusters used to construct the convex hull and the maximum number of atoms included in the derivative structures.

Note that the obtained convex hull is a grand-canonical one.
For alloys, a canonical convex hull including phase separation states must be considered. 
This corresponds to a cross-section surface for a constant composition.
Nevertheless, we only have to consider grand-canonical vertices since canonical vertices belong to either grand canonical vertices on the constant composition or the phase separation states of vertices on the other compositions.

\section{Convex hulls for simple lattices}
\label{special:convex_hull_simple_lattice}
In this section, convex hulls for bcc, fcc, hcp and sc lattices are shown.
As described above, the accuracy of the convex hull is determined by both the number of clusters and the maximum number of atoms included in the derivative structures.
Here we consider the correlation functions only for point and pair clusters up to the fifth nearest-neighbor (NN) pair to construct the convex hulls.
Derivative structures with up to 16 atoms are employed to construct the convex hulls.
They are composed of 154158, 154158, 90863 and 177676 structures for the bcc, fcc, hcp and sc lattices, respectively, including end members and structures where atomic species are exchanged.
Then, the correlation functions of all derivative structures are calculated.
The search of the derivative structures and the computation of the correlation functions are carried out using the {\sc clupan} code\cite{clupan,ame_feature:Seko,casp:seko}.

\begin{figure}[tbp]
\begin{center}
\includegraphics[width=\linewidth]{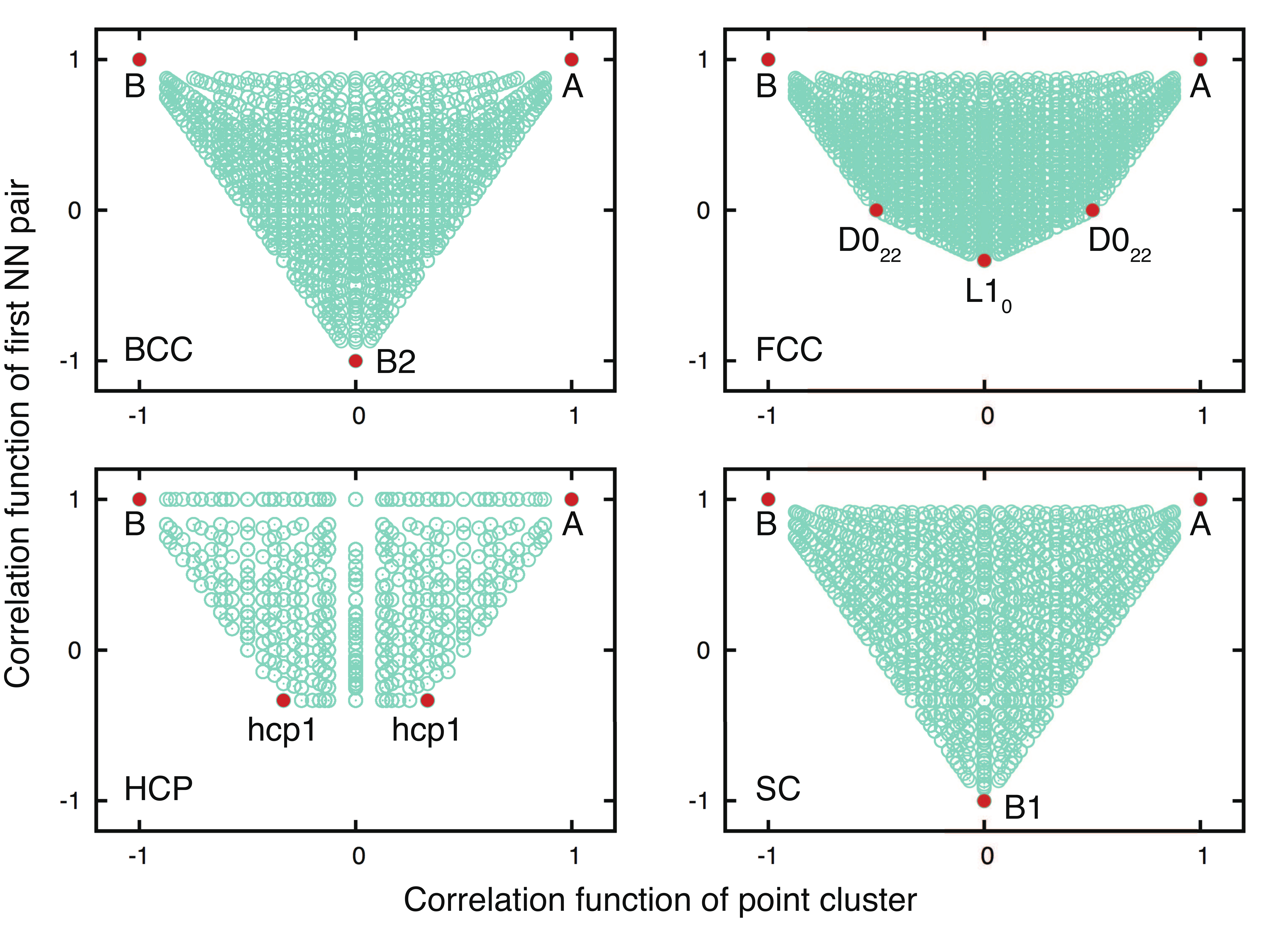} 
\caption{
Distributions of derivative structures in two-dimensional space of correlation functions of the point and NN pair for bcc, fcc, hcp and sc lattices.
The vertex structures are shown by red closed circles.
Crystal structures of the vertex structures that are not categorized into any prototype structure are shown in Fig. \ref{special:vertex_structure}.
For the hcp lattice, the NN pair corresponds to that on (0001) plane.
}
\label{special:distribution_twodim}
\end{center}
\end{figure}

Figure \ref{special:distribution_twodim} shows distributions of the correlation functions of the point and NN pair for all the derivative structures.
The vertex structures for the point and NN pair are also shown.
When there are multiple structures with the same correlation functions, only the structure expressed by the smallest number of atoms is shown.
For example, the L1$_0$ and `NbP' structures have the same correlation functions for the point and NN pair, hence only the L1$_0$ structure is shown in Fig. \ref{special:distribution_twodim}.
As can be seen in Fig. \ref{special:distribution_twodim}, the convex hulls for the bcc and sc lattices are triangles because they do not exhibit the frustration effect for the NN pair. 
Therefore, except for the end members, only one structure is the vertex structure. 
The vertex structures for the bcc and sc lattices are the B2 and B1 structures, respectively.
On the other hand, the convex hull for the fcc lattice is not a triangle but is symmetric with respect to the correlation function of the point cluster owing to the frustration effect of the NN pair.
Therefore, there are five vertex structures, which are the L1$_0$ and D0$_{22}$ structures in addition to the end members.

\begin{figure}[tbp]
\begin{center}
\includegraphics[width=\linewidth]{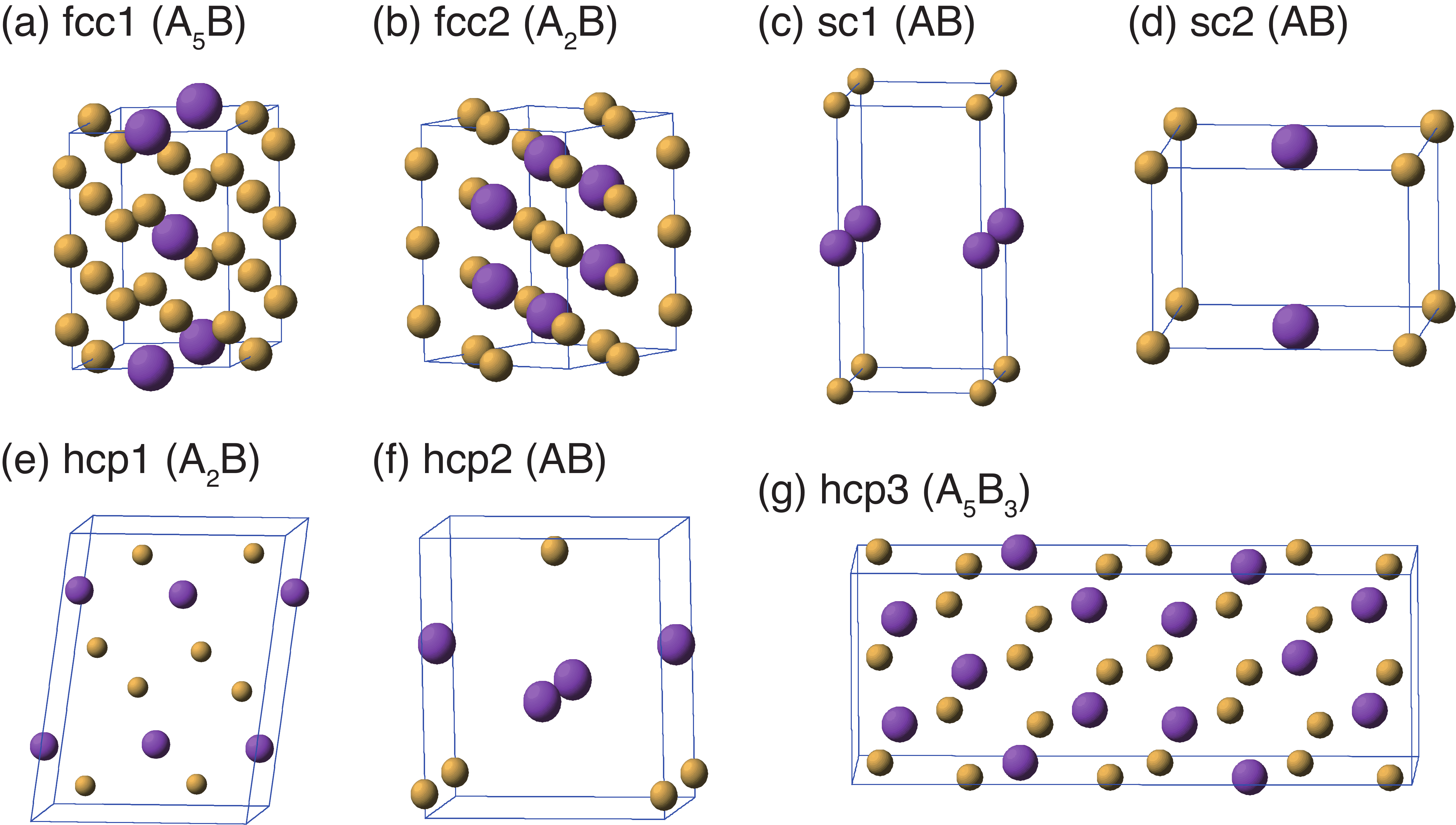} 
\caption{
Illustration of crystal structures of vertex structures that are not categorized into any prototype structure.
}
\label{special:vertex_structure}
\end{center}
\end{figure}

The vertex structures for the point, first NN and second NN pairs are listed in Table \ref{special:structure_2ndNN}.
Crystal structures of the vertex structures that do not correspond to any prototype structure are shown in Fig. \ref{special:vertex_structure}.
Some of the convex hulls have already been derived analytically in the literature\cite{CE3,kanamori1966,kaburagi1975method,kanamori1977conditions,lipkin1988ground,PhysRevB.27.3018,richards1971pairwise,allen1972ground,allen1973ground,cahn1979ground,sanchez1981structure,kudo1976method}.
The vertex structures for the fcc lattice with the first and second NN pair interactions are exactly the same as those of the analytical convex hull\cite{CE3,kanamori1966}.

\begin{table*}[tbp]
\caption{
Vertex structures in three-dimensional space of correlation functions of the point, first and second NN pairs for the bcc, fcc, hcp and sc lattices.
When there are multiple structures on a vertex, the structure described by the smallest number of atoms is shown.
V.S. (1st NN) stands for vertex structures only for the first NN pair.
}
\label{special:structure_2ndNN}
\begin{ruledtabular}
\begin{tabular}{lll}
& Up to first NN pair & Up to second NN pair \\
\hline
bcc & A, B, B2 & V.S.(1st NN) $+$ B32, D0$_3$$\times 2$\\
fcc & A, B, D0$_{22}$$\times 2$, L1$_0$ & V.S.(1st NN) $+$ L1$_1$, `NbP', `MoPt$_2$'$\times 2$, L1$_2$$\times 2$, A$_5$B(fcc1)$\times 2$, A$_2$B(fcc2)$\times 2$ \\
hcp & A, B, A$_2$B(hcp1)$\times 2$ & V.S.(1st NN) $+$ B$_h$, B19, D0$_a$$\times 2$, AB(hcp2), A$_3$B$_5$(hcp3) \\
sc  & A, B, B1 & V.S.(1st NN) $+$ `AuSb$_3$'$\times 2$, AB(sc1), AB(sc2) 
\end{tabular}
\end{ruledtabular}
\end{table*}

\begin{figure}[tbp]
\begin{center}
\includegraphics[width=\linewidth]{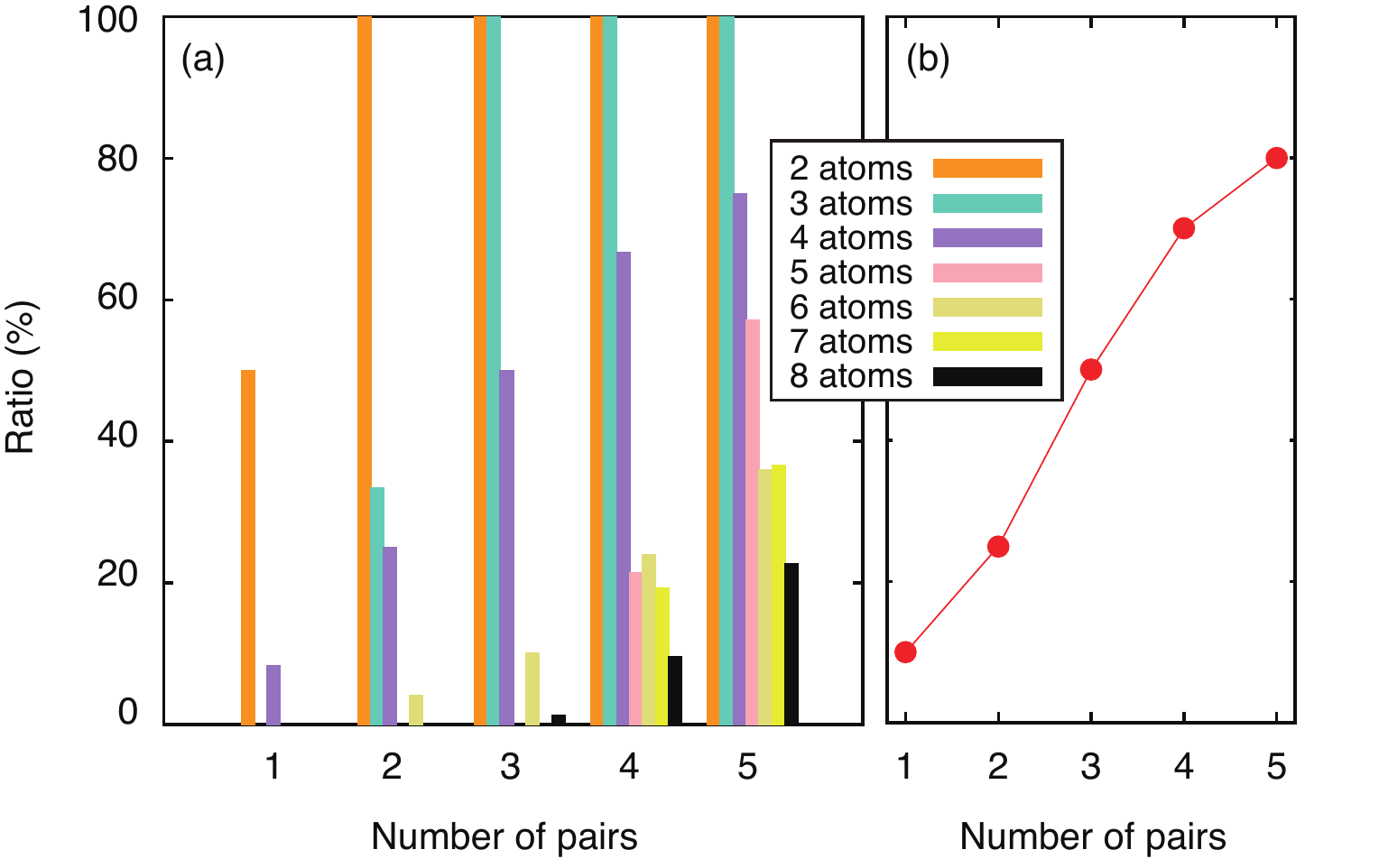} 
\caption{
(a) Ratio of the number of fcc derivative structures corresponding to the vertex structures to the total number of fcc derivative structures.
(b) Ratio of the number of fcc prototype structures corresponding to the vertex structures to the total number of fcc prototype structures in the ICSD database.
}
\label{special:n_structure_fcc}
\end{center}
\end{figure}

Table \ref{special:n_special} shows the dependence of the number of vertex structures on the number of pairs.
Naturally, the number of vertex structures increases with the number of interactions. 
The ratio of the number of derivative structures corresponding to the vertex structures to the total number of derivative structures for the fcc lattice is shown in Fig. \ref{special:n_structure_fcc} (a).
As can be seen in Fig. \ref{special:n_structure_fcc} (a), the derivative structures expressed by a small number of atoms are likely to be the vertex structures.
The ratio for the derivative structures with a smaller number of atoms is higher than that for the derivative structures with a larger number of atoms.
In particular, all the derivative structures expressed by two or three atoms correspond to the vertex structures by considering pairs up to the third NN pair.
As a result, the majority of vertex structures correspond to derivative structures expressed by a small number of atoms when considering a small number of short-range interactions.

\begin{table}[tbp]
\caption{
Dependence of the number of vertex structures on the number of pairs for bcc, fcc, hcp and sc lattices.
The total number of derivative structures with up to 16 atoms is also shown.
The hcp lattice has six symmetrically-independent pairs up to 4th NN.
}
\label{special:n_special}
\begin{ruledtabular}
\begin{tabular}{lcccc}
& \multicolumn{4}{c}{Number of structures} \\
& bcc & fcc & hcp & sc \\
\hline
Vertex structures & & & & \\
Up to 1st NN & 3  & 5  & 11   & 3  \\
Up to 2nd NN & 6  & 15  & 25   & 9  \\
Up to 3rd NN & 60 & 68  & 93   & 19 \\
Up to 4th NN & 230 & 646 & 2459  & 291 \\
%Up to 5th NN & 619 & 2224 & ***  & 1037 \\
\hline
Derivative structures  & 154158  & 154158 & 90863 & 177676 
\end{tabular}
\end{ruledtabular}
\end{table}

Figure \ref{special:n_structure_fcc} (b) shows the ratio of the number of fcc prototype structures that are vertex structures to the total number of fcc prototype structures included in the Inorganic Crystal Structure Database (ICSD)\cite{ICSD}.
Here the prototype structures expressed by up to 8 atoms are considered.
Simply by considering pairs up to the fifth NN pair, 80\% of the prototype structures are included in the vertex structures.
Sixteen prototype structures are included in the vertex structures.
They are
B11, C6, C11$_b$, D0$_{22}$, D1$_a$, L1$_0$, L1$_1$, L1$_2$
`NbP', `MoPt$_2$', `Ca$_7$Ge', `Au$_5$Mn$_2$', `Ga$_3$Pt$_5$', `Nb$_3$Au$_2$', `UGe$_2$' and `ZrSi$_2$' structures, while only four prototype structures are not included in the vertex structures: D0$_{23}$, `Pd$_5$Ti$_3$', `Cu$_4$Ti$_3$' and `Al$_3$Os$_2$' structures.
Except for the D0$_{23}$ structure, they are seldom found in the ICSD.
These results imply that a small number of short-range pair interactions are dominant in determining the ground-state structures.

\section{Applications}
\subsection{Cluster expansion of Ag-Au binary alloy}
We examine the accuracy of the vertex structures by applying them to Ag-Au binary alloy.
Here we compute the formation energies of all the derivative structures using the CE method instead of the DFT calculation since the DFT calculation for all derivative structures is computationally prohibitive.
%Once the ECIs are estimated, the energies of all derivative structures are easily computed by the CE.
The ground-state structures obtained from the formation energies of all derivative structures are regarded as the true ground-state structures.
The ground-state structures obtained only from the vertex structures are then compared with the true ground-state structures. 

To construct the CE, DFT calculations for 140 input structures are first performed by the projector augmented-wave (PAW) method\cite{PAW1,PAW2} within the Perdew-Burke-Ernzerhof exchange-correlation functional\cite{GGA:PBE96} as implemented in the \textsc{VASP} code\cite{VASP1,VASP2}.
The total energies converge to less than 10$^{-3}$ meV/supercell.
The atomic positions and lattice constants are relaxed until the residual forces become less than $10^{-2}$ eV$/$\AA.
Using a set of DFT energies, ECIs are estimated using a least-squares fitting without a regularization term.
An optimized set of clusters with the minimized leave-one-out cross validation (CV) score is selected by a genetic algorithm\cite{geneAlgo1,geneAlgo2}.
The optimized set of clusters is composed of the empty, point, three pairs, two triangles and three quadruplets, and has a CV score of 0.5 meV/atom.
By computing the formation energies of all derivative structures and vertex structures from the obtained ECIs, the ground-state structures are estimated.

\begin{figure}[tbp]
\begin{center}
\includegraphics[width=\linewidth]{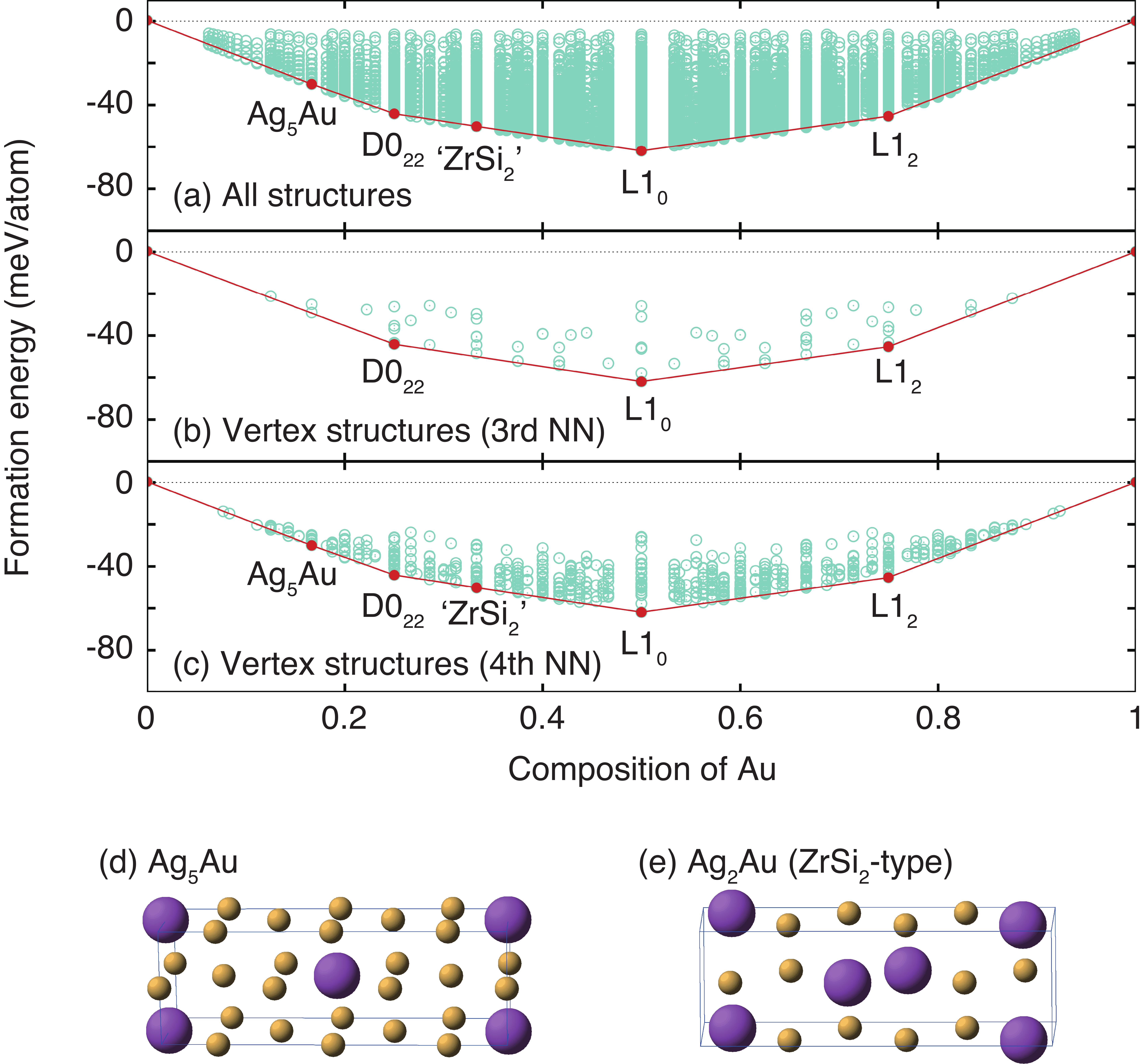} 
\caption{
Formation energies computed by the CE method for (a) all derivative structures, (b) vertex structures for pairs up to the third NN and (c) vertex structures for pairs up to the fourth NN in Ag-Au alloy.
%The red lines show the convex hulls of the formation energies, which correspond to the ground states.
The red lines show the energies of the ground states structures.
The crystal structures of (d) Ag$_5$Au and (e) Ag$_2$Au are also shown.
Small and large spheres represent Ag and Au, respectively.
}
\label{special:gs_Ag-Au}
\end{center}
\end{figure}

Figure \ref{special:gs_Ag-Au} (a) shows the formation energies of all 154158 derivative structures.
%The ground state corresponds to the convex hull of the formation energies.
The CE derives five stable intermetallic compounds: Ag$_5$Au, Ag$_3$Au (D0$_{22}$), Ag$_2$Au (`ZrSi$_2$'), AgAu (L1$_0$) and AgAu$_3$ (L1$_2$).
In the experimental phase diagram of Ag-Au alloy\cite{hassam1988experimental}, only the fcc solid solution phase was reported.
In a paper by Ozoli{\c{n}}{\v{s}} et al., using a combination of DFT calculations and the CE method, the ground-state structures were predicted to be the L1$_2$ structure for the Au compositions of 0.25 and 0.75 and the L1$_0$ structure for the Au composition of 0.5\cite{ozolicnvs1998cu}.
These structures are partially consistent with our prediction.

Figures \ref{special:gs_Ag-Au} (b) and (c) show the formation energies of vertex structures obtained from pairs up to the third NN and fourth NN, which are composed of 68 and 646 vertex structures, respectively.
From the calculation of the formation energies of vertex structures up to the third NN, three of the true ground-state structures, Ag$_3$Au (D0$_{22}$), AgAu (L1$_0$) and AgAu$_3$ (L1$_2$), are predicted since the `ZrSi$_2$' and Ag$_5$Au structures are not included in the vertex structures.
However, the energies of the ground-state structures obtained from the vertex structures are almost the same as those of the true ground-state structures.
On the other hand, the ground-state structures obtained from the vertex structures for pairs up to the fourth NN pair agree with all the true ground-state structures even though not all the selected ECIs are considered.
The true ground-state structures are efficiently obtained by considering only 646 structures instead of all the derivative structures.

\subsection{Nine metallic alloys}
Next we evaluate the most stable structure in alloys by performing the DFT calculation for only the vertex structures.
We employ nine intermetallic compounds: CuAu, CuAg, CuPd, AuAg, AuPd, AgPd, MoTa, MoW and TaW.
The vertex structures with the composition of 0.5 are selected from a set of grand-canonical vertex structures obtained from pairs up to the fourth NN for the bcc, fcc and hcp lattices. 
The total number of vertex structures with the composition of 0.5 is 56.
The computational detail of the DFT calculation is the same as that for the CE for Ag-Au alloy.

\begin{figure}[tbp]
\begin{center}
\includegraphics[width=0.9\linewidth]{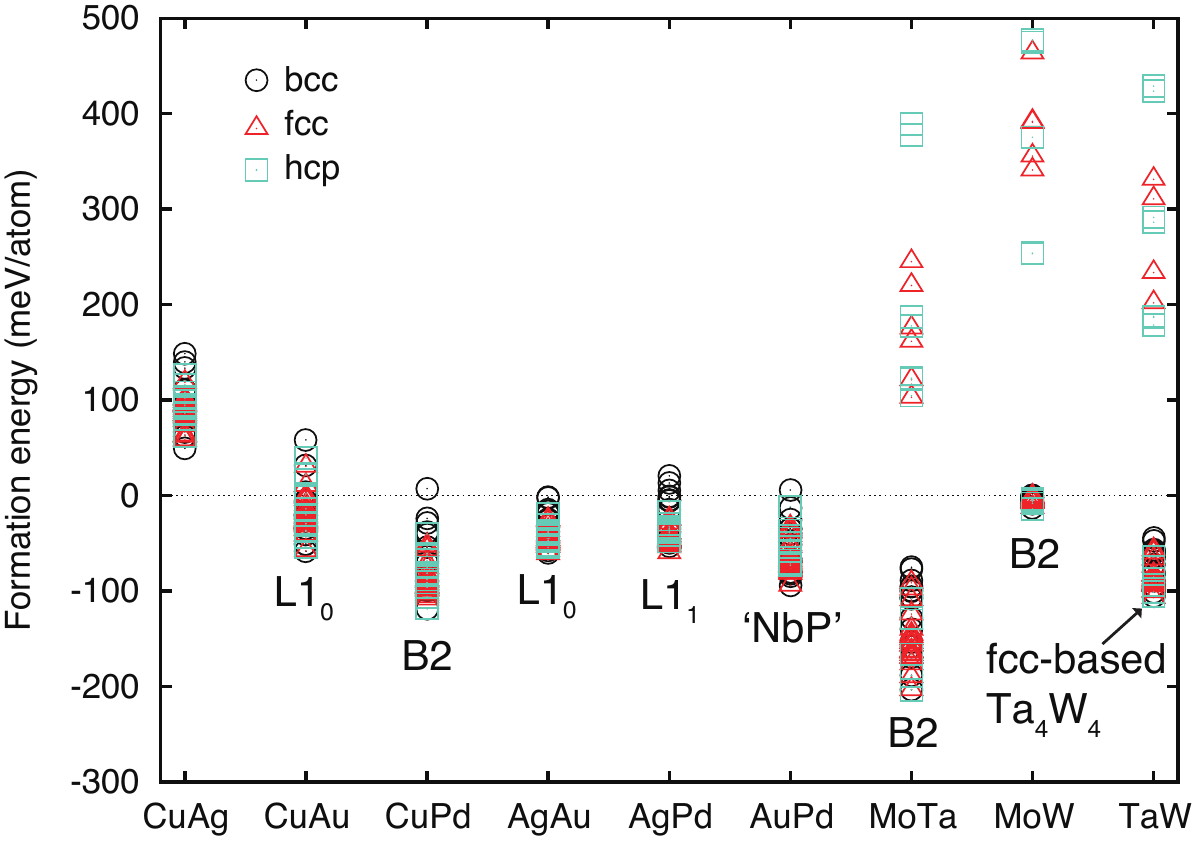} 
\caption{
Formation energies of vertex structures for CuAu, CuAg, CuPd, AuAg, AuPd, AgPd, MoTa, MoW and TaW.
The energies of the structures after performing structure optimization starting from the vertex structures for pairs up to the fourth NN for the bcc, fcc and hcp lattices are shown by black open circles, red open triangles and blue open squares, respectively.
}
\label{special:energy_alloys}
\end{center}
\end{figure}

Figure \ref{special:energy_alloys} shows the formation energies of the vertex structures for the bcc, fcc and hcp lattices in the nine compounds.
For CuAg, the formation energies of all vertex structures are positive, hence the phase-separated state is the most stable.
This is consistent with the experimental phase diagram\cite{Ag-Cu_phase_diagram}.
On the other hand, the formation energies of the most stable structure are negative for the other eight compounds.
For CuAu and CuPd, the predicted ground states are the L1$_0$ and B2 structures, respectively, which agree with the experimentally reported structures\cite{Cu-Au_phase_diagram,Cu-Pd_phase_diagram} and theoretical ground-state structures\cite{ozolicnvs1998cu,trimarchi2008finding}.
For AgAu, AgPd, AuPd, MoTa, MoW and TaW, the predicted ground-state structures correspond to L1$_0$, L1$_1$, `NbP', B2, B2 and fcc-based Ta$_4$W$_4$ structures, respectively.
Although the existence of ordered structures is experimentally unknown and an fcc or bcc solid solution has been reported to be the stable phase for the entire range of compositions for the six alloys\cite{hassam1988experimental,Ag-Pd_phase_diagram,Au-Pd_phase_diagram,Mo-Ta_phase_diagram,Mo-W_phase_diagram,Ta-W_phase_diagram}, theoretical predictions of the ground-state structures have been reported for four of the six alloys. 
For AgAu, AgPd, AuPd and MoTa, the ground-state structures were predicted to be L1$_0$\cite{ozolicnvs1998cu}, L1$_1$\cite{PhysRevLett.87.165502}, `NbP'\cite{PhysRevB.74.035108} and B2 structures\cite{PhysRevB.70.155108}, respectively, which are in good agreement with our prediction.

\subsection{Ionic compound}
Finally, we show an application to an ionic system with a complex lattice.
We adopt the heterovalent cation configuration in MgAl$_2$O$_4$ spinel as the complex system.
The validity of our procedure is demonstrated by comparing the energy distribution obtained from all derivative structures with a fixed number of atoms and that obtained from the vertex structures.
We focus on the energy distribution because the vertex structures are likely to include not only the ground-state structure but also the structure with the highest energy.

A spinel compound with cations A and B and anion C has the general formula AB$_2$C$_4$, where the anions form a nearly fcc sublattice. 
The spinel structure has two types of cation site, namely tetrahedral fourfold-coordinated and octahedral sixfold-coordinated sites. 
The number of octahedral sites is double the number of tetrahedral sites.
When all the tetrahedral sites are occupied by cation A, the spinel is called ``normal'', and when all the tetrahedral sites are occupied by cation B, the spinel is called ``inverse''. 

Although it is necessary to consider a large number of long-range ECIs in heterovalent ionic systems\cite{Seko_longrange_Jphys}, we explore vertex structures using a small number of short-range pairs.
Vertex structures are obtained from derivative structures with up to 18 cations (42 atoms) for the cation lattice of the spinel structure.
All the derivative structures with the composition of MgAl$_2$O$_4$ are considered here, hence no phase separation states between two different compositions are considered.
Therefore, a canonical convex hull excluding the phase separation states is obtained.
Table \ref{special:n_special_spinel} shows the number of vertex structures on the canonical convex hull for the cation lattice of the spinel structure.

\begin{table}[tbp]
\caption{
Number of vertex structures for the cation lattice of the spinel structure compared with the number of derivative structures with up to 18 cations (42 atoms).
}
\label{special:n_special_spinel}
\begin{ruledtabular}
\begin{tabular}{lc}
& Number of structures \\
\hline
Vertex structures & \\
Up to 2nd NN & 6 \\
Up to 3rd NN & 17 \\
Up to 4th NN & 39 \\
Up to 5th NN & 117 \\
\hline
Derivative structures & 2366
\end{tabular}
\end{ruledtabular}
\end{table}

\begin{figure}[tbp]
\begin{center}
\includegraphics[width=\linewidth]{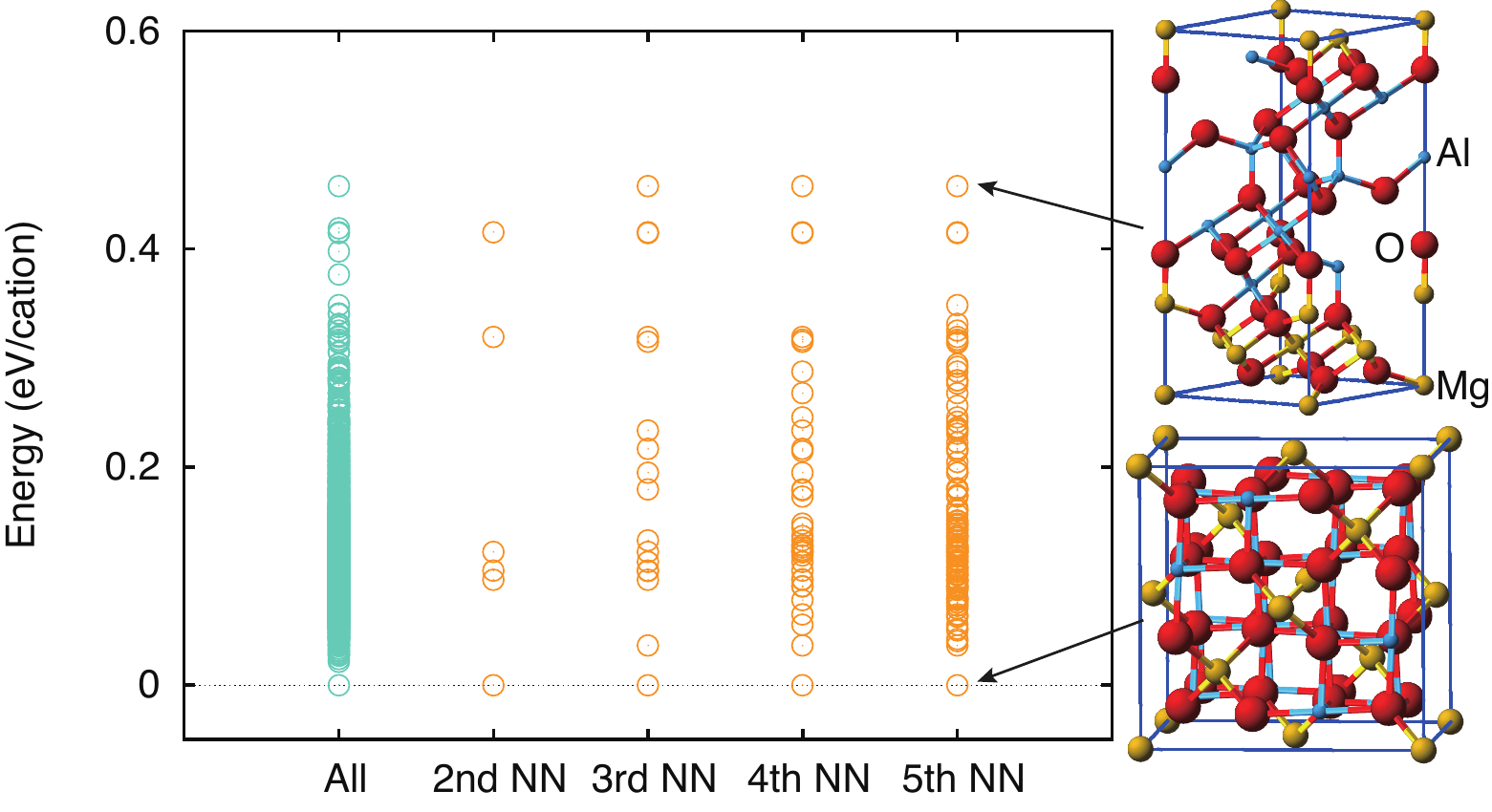} 
\caption{
Energy distribution of the vertex structures obtained from pairs up to the second, third, fourth and fifth NN in MgAl$_2$O$_4$. 
The energy distribution of all derivative structures is also shown by blue open circles.
The energy is measured from that of the normal spinel.
Crystal structures for the ground state and highest-energy structures are also shown.
}
\label{special:energy_MgAl2O4}
\end{center}
\end{figure}

Figure \ref{special:energy_MgAl2O4} shows the energy distribution of the vertex structures along with that of all derivative structures in MgAl$_2$O$_4$ spinel.
Since the ground-state structure of the normal spinel is included in the vertex structures by considering only the first NN pair, the ground-state structure is correctly obtained by computing the energies of a small number of vertex structures. 
On the other hand, the structure with the highest energy is obtained after including the third NN pair.
Nevertheless, the ground state and highest-energy structures are correctly predicted by computing the energies of only 17 vertex structures for pairs up to the third NN.

It appears strange that the ground-state structure can be estimated from the calculation for the vertex structures obtained by considering only a few short-range interactions even in an ionic multicomponent system with the configurations of heterovalent ions.
However, this is ascribed to the fact that only the short-range ECIs are required to accurately express the energy for a short-period structure, while the long-range ECIs are essential only for long-period structures\cite{Seko_longrange_Jphys}.
Therefore, as long as the ground-state structures are searched for among a set of structures expressed by a small number of atoms, it is practically acceptable to consider only short-range interactions.
This may be justified by the empirical fact that most ionic compounds take simple crystal structures despite interatomic interactions being long-range.

\section{Conclusion}
An efficient approach to determining the ground-state structures in alloys has been demonstrated. 
In this approach, the computation of energy is only required for vertex structures of the configurational polyhedron describing the range of correlation functions.
Here the configurational polyhedron is approximated by the convex hull estimated from the correlation functions of all possible structures with up to a fixed number of atoms using an efficient numerical algorithm. 
In this study, we have clarified the vertex structures, which are obtained from a small number of short-range pair interactions for four types of simple lattice, namely, bcc, fcc, hcp and sc lattices. 
By comparing the vertex structures with the observed prototype structures, most of the prototype structures are found to correspond to the vertex structures. 
This implies that a small number of short-range pair interactions are dominant in determining the ground-state structures. 
We then applied the method to three types of system as follows.
(1) A ground state search for Ag-Au alloy using a combination of DFT calculation and the CE method.
(2) A ground state search for nine intermetallic compounds: CuAu, CuAg, CuPd, AuAg, AuPd, AgPd, MoTa, MoW and TaW. 
(3) The energy distribution in MgAl$_2$O$_4$ spinel with different cation arrangements, where the crystal structure is more complex than that of close-packed structures and long-range electrostatic interactions must be considered. 
In these applications, the ground-state structures were successfully found from vertex structures with only a small number of pair interactions. 
These results indicate that our procedure can be applied to the efficient exploration of ground-state structures in a wide range of systems.

\begin{acknowledgments}
This study was supported by a Grant-in-Aid for Scientific Research on Innovative Areas ``Nano Informatics'' (grant number 25106005) from Japan Society for the Promotion of Science (JSPS).
\end{acknowledgments}

\bibliography{special}
\end{document}